%% file: output.tex
\documentclass{article}

\usepackage[margin=1in]{geometry}
\usepackage{indentfirst}

\usepackage[utf8]{inputenc} % allow utf-8 input
\usepackage[T1]{fontenc}    % use 8-bit T1 fonts
\usepackage{hyperref}       % hyperlinks
\usepackage{url}            % simple URL typesetting
\usepackage{booktabs}       % professional-quality tables
\usepackage{amsfonts}       % blackboard math symbols
\usepackage{nicefrac}       % compact symbols for 1/2, etc.
\usepackage{microtype}      % microtypography

\usepackage{bbm}
\usepackage{amsfonts}
\usepackage{amsmath,amsthm}           
\usepackage{mathtools}    
\usepackage{amsopn}
\usepackage{amssymb}
\usepackage{bm}
\usepackage{multirow}
\usepackage{graphicx}
\usepackage{subfigure}
\usepackage{adjustbox}
\usepackage{xcolor}
\usepackage[vlined,linesnumbered,ruled]{algorithm2e}
\usepackage{biblatex}
\usepackage{tabularx}
\usepackage{ragged2e}
\usepackage{booktabs}

\newtheorem{assumption}{Assumption}
\newtheorem{theorem}{Theorem}

\newtheorem{proposition}{Proposition}

\newtheorem{obser}{Observation}

\allowdisplaybreaks

\DeclareMathOperator*{\argmax}{\arg\!\max}

\newcolumntype{L}{>{\RaggedRight\hangafter=1\hangindent=0em}X}

\allowdisplaybreaks

\title{Full Stage Learning to Rank: A Unified Framework for Multi-Stage Systems}
\author{
Kai Zheng\\
Kuaishou Inc. \\
\texttt{zhengk92@gmail.com}
\and
Haijun Zhao\\
Sun Yat-Sen University \\
\texttt{zhaohj23@mail2.sysu.edu.cn}
\and
Rui Huang\\
Kuaishou Inc.\\
\texttt{huangrui06@kuaishou.com}
\and 
Beichuan Zhang\\
Kuaishou Inc.\\
\texttt{zhangbeichuan@kuaishou.com}
\and
Na Mou\\
Kuaishou Inc. \\
\texttt{mouna@kuaishou.com}
\and
Yanan Niu\\
Kuaishou Inc. \\
\texttt{niuyanan@kuaishou.com}
\and
Yang Song\\
Kuaishou Inc. \\
\texttt{yangsong@kuaishou.com}
\and
Hongning Wang\\
Tsinghua University \\
\texttt{wang.hongn@gmail.com}
\and
Kun Gai\\
Independent \\
\texttt{gai.kun@qq.com}
\and
}
\date{}

\begin{document}
\maketitle

\begin{abstract}
  The Probability Ranking Principle (PRP) has been considered as the foundational standard in the design of information retrieval (IR) systems. The principle requires an IR module's returned list of results to be ranked with respect to the underlying user interests, so as to maximize the results' utility. 
  Nevertheless, we point out that it is inappropriate to indiscriminately apply PRP through every stage of a contemporary IR system. Such systems contain multiple stages (e.g., retrieval, pre-ranking, ranking, and re-ranking stages, as examined in this paper).  The \emph{selection bias} inherent in the model of each stage significantly influences the results that are ultimately presented to users.
  To address this issue, we propose an improved ranking principle for multi-stage systems, namely the Generalized Probability Ranking Principle (GPRP), to emphasize both the selection bias in each stage of the system pipeline as well as the underlying interest of users. 
 We realize GPRP via a unified algorithmic framework named Full Stage Learning to Rank. Our core idea is to first estimate the selection bias in the subsequent stages and then learn a ranking model that best complies with the downstream modules' selection bias so as to deliver its top ranked results to the final ranked list in the system's output.
 We performed extensive experiment evaluations of our developed Full Stage Learning to Rank solution, using both simulations and online A/B tests in one of the leading short-video recommendation platforms. The algorithm is proved to be effective in both retrieval and ranking stages. Since deployed, the algorithm has brought consistent and significant performance gain to the platform.
\end{abstract}

\input{introduction.tex}
\input{related.tex}
\input{preliminary.tex}
\input{gprp.tex}

\input{method.tex}
\input{assum_verify.tex}
\input{practice.tex}
\input{experiment.tex}
\input{conclusion.tex}

% \begin{acks}
% \end{acks}

\newpage
\printbibliography

\newpage
\appendix
\input{appendix.tex}
\end{document}

%% file: introduction.tex
\section{Introduction}
\label{sec:intro}
Information retrieval (IR) systems widely influence numerous aspects of our daily lives, from information search to entertainment choices and travel decisions. The significant impact has gathered extensive research interest and gained substantial attention since its inception \cite{singhal2001modern}. Central to this research area is the precise inference of users' underlying information needs, so as to present the most relevant information to the users, a concept that proved to be optimal by the Probability Ranking Principle (PRP) \cite{robertson1977probability}.

Early year's IR research only concerned single-stage systems, for example using BM25 \cite{robertson2009probabilistic} or a logistic regression model \cite{gey1994inferring} to rank results against a given query or user. Propelled by the emergency of new demands and development of technologies, modern IR systems nowadays are equipped with multiple stages, such as retrieval and ranking stages in YouTube \cite{youtubeDNN}, the recall, first round ranking, and second round ranking stages in Yahoo search \cite{yin2016ranking}, and the retrieval, pre-ranking, ranking and re-ranking stages in KuaiShou \cite{wang2020cold}. 
No matter it is manually crafted rules, such as the BM25 scoring function, or data-driven rankers, such as ranking models \cite{liu2009learning}, in each component of these multi-stage IR systems, PRP is still the design principle, i.e., each stage is supposed to present the most relevant results for the next stage's further processing. 

%No matter in one stage or multi-stage systems, there exists selection bias introduced by current system, that is our training data is collected after the filtration of current recommendation system and the model learned based on training data still constitute part of the system. This phenomenon causes additional difficulty in learning users' underlying interests, since some interests may never be observed, let alone learning them. Thus, there is a very important line of research studying how to eliminate selection bias and learn users' unbiased interests \cite{chen2019top,joachims2017unbiased,hu2019unbiased,saito2020unbiased}. Note the principle behind unbiased learning when deployed in online recommendation systems is still PRP, i.e. recommending items based on users' unbiased estimated interests. 

However, the optimality of PRP strongly depends on the assumption that the top-ranked result by the algorithm will eventually be presented to the end users. This is only true in a single-stage IR system, but no longer holds in multi-stage systems. For example, at the retrieval stage (typically an early stage), even if we know the underlying preferences of users on each result in the whole candidate set and return the most relevant ones to the later stage, there is a possibility that some or even all of them are filtrated eventually because the \emph{selection bias} of subsequent stages can deviate from user's underlying interests and promote sub-optimal results along the way until reaching the end users \footnote{In a commercial system, there can be tens of different ranking algorithms employed in each stage of the pipeline, making the selection bias inevitable.}.

The fundamental reason for the \emph{selection bias} resides in the misalignment between how the ranking models in each stage are learned and how they are used in a multi-stage system. Typically, the training data for each stage is constructed by treating the results from this stage and finally preferred by the end users as positive and the rest from this stage as negative. Sub-sampling can be devised when the stage returns a large number of results (e.g., more than thousands), but the end users can only interact with a few (e.g., less than ten). 
However, relevant results from the intermediate stage but eventually filtered by later stages cannot be differentiated from those truly irrelevant ones, and are often mistakenly treated as negatives. 
This phenomenon is analogous to position bias in user feedback \cite{joachims2017accurately,craswell2008experimental}, where non-clicked results do not suggest irrelevance, as they could also not be examined by a user. In a multi-stage system, the selection bias is introduced by the subsequent components in an IR pipeline, which is more implicit and dynamic. Hence, those well-known solutions for correcting position bias \cite{chen2019top,joachims2017unbiased,hu2019unbiased,saito2020unbiased} do not apply to this problem.

Moreover, the operation and management of multi-stage systems in industry practice further exacerbates the selection bias. Due to the system's high complexity, each stage or even a particular algorithm at a single stage is often managed by a different team. When improving an algorithm or choosing which algorithm to deploy, each team can only access and control data at its specific stage. However, these intermediate algorithmic decisions are only informed by online A/B tests on the final results. 

%Similarly, PRP is also sub-optimal in non-final stages of mutli-stage recommendation system because of the discrepancy of preference between subsequent stages and users, except in the last stage where we can show items directly to users. In practice, usually we can only handle the algorithmic model in one stage of a multi-stage recommendation system, because of the principle of online A/B testing or the fact that different teams handle different stages and we only have permission to modify a certain stage. Therefore, traditional PRP may not be suitable in modern multi-stage recommendation systems, which could lead to sub-optimal performance. 

In this paper, we propose an improved version of PRP named the Generalized Probability Ranking Principle (GPRP) for multi-stage IR systems. GPRP extends PRP by explicitly modeling the ranking preferences from both the end users and subsequent stages. GPRP degrades to PRP when there only one stage is present or it functions at the final stage of the system. At non-final stages, the expected ranking utility achieved by GPRP is always an upper bound of the expected utility achieved by PRP, which can also be proved to be a max-min approximation of GPRP. However, it is difficult to precisely fulfill GPRP in practice due to computational complexity. Under mild assumptions about a multi-stage IR system, 
%which can be approximately verified in our online platform, one of largest short-video platform in the world, 
we develop an efficient and effective algorithmic framework named Full Stage Learning to Rank (FS-LTR) to realize GPRP approximately. The core idea of FS-LTR is to define preferential treatments on the exposure data collected in each stage of the system as well as users' feedback for the final ranking and fit an LTR algorithm with the preferential labels to estimate the most effective ranking for each particular stage. FS-LTR can be seamlessly applied to any stage in the entire pipeline. Extensive experiments on both a simulated environment and online A/B testing in one of the world's largest short-video recommendation platforms validate the effectiveness of our framework. 
To the best of our knowledge, we are the first to study ranking principles for multi-stage IR systems holistically and to design a universally efficient algorithmic framework with 
theoretical foundations. 

%In summary, our main contributions are:
%\begin{itemize}
%    \item[1] We formalize the process of multi-stage recommendation systems, and take a small step towards understanding the behavior such complex systems;
%    \item[2] We propose GPRP which is an improved version of PRP in multi-stage recommendation systems and formally study the relation between GPRP and PRP;
%    \item[3] Under mild assumptions about multi-stage recommendation systems, we design a simple but efficient and effective algorithmic framework which can be proved to achieve GPRP approximately, and it can be used either in retrieval stage or any other stage of the system;
%    \item[4] We give our practical experience about the implementation of FS-LTR in real-world multi-stage recommendation system, including data collection, training and serving; 
%    \item[5] Both simulated experiments and online A/B testing demonstrate the effectiveness of our algorithmic framework. 
%\end{itemize}

%% file: related.tex
\section{Related Work}
\label{sec:related}
\textbf{Single Stage Optimization:} Most of existing work consider optimizing either retrieval or general ranking model. For example, there are various different retrieval algorithms including basic dual-encoder model \cite{youtubeDNN,MIND_model,comirec}, tree-based deep models \cite{TDM}. General ranking can be more carefully divided to three (or even more) stages, including pre-ranking \cite{wang2020cold,liu2017cascade}, ranking \cite{youtubeDNN,zhou2018deep,zhou2019deep,pi2020search,chang2023pepnet,chang2023twin}, and re-ranking \cite{burges2010ranknet,pei2019personalized,feng2021grn,jia2021pairrank}. The difference between models in the three stages mainly lies in their model capacity and training method, either in a pointwise, pairwise, or listwise manner.

\noindent \textbf{Multi-stage Joint Optimization:} Different from optimizing any single stage of a multi-stage system, this line of research considers how to optimize them jointly, either taking derivative with respect to each stage's model parameters \cite{gallagher2019joint} or collecting many samples in each stage to train each ranker alternatively \cite{qin2022rankflow}. However,  it's hard to fulfill joint optimization of all stages at the same time in a practical system, since different stages are often managed by different teams. Besides, these methods seem hard to be deployed in real systems as far as we know, mainly because of highly complex real system or computational overhead.

\noindent \textbf{Unbiased Learning:} The goal of unbiased learning is to eliminate the exposure bias introduced by training data collected from recommendation systems and learn unbiased ground-truth interests of users. one main idea of unbiased learning is based on the Inverse Propensity Score (IPS) estimator \cite{rubin1974estimating}, which is used either at the retrieval stage \cite{chen2019top} or re-ranking stage \cite{joachims2017unbiased,saito2020unbiased,oosterhuis2020policy,ovaisi2020correcting,ovaisi2021propensity}.

%% file: preliminary.tex
\section{Preliminary of Multi-Stage Systems}
\label{sec:preliminary}
Without loss of generality, we consider a multi-stage system with four stages: retrieval (stage 1, from all candidates to $c_1$ items), pre-ranking (stage 2, from $c_1$ items to $c_2$ items), ranking (stage 3, from $c_2$ items to $c_3$ items), and re-ranking (stage 4, from $c_3$ items to $c_4$ items), where $c_1, c_2, c_3, c_4$ are predefined values like 10000/500/50/6 respectively. 
% or random variables around predefined numbers where randomness is introduced by strategies in systems and two numbers in each bracket represent the input candidate number and output item number in each stage given a request from the user to the system. 
Note the output of one stage is exactly the input candidate set of the next stage. Finally, $c_4$ items outputted by the re-ranking stage are exposed to a user and receive corresponding user feedback.
%, then a new request to the system begins until the user leaves the system. 
Given a candidate item set $A$, a real-valued function $f(u, v)$ and a positive integer $c$, the $\textbf{Topk}(A, f(u, v), c)$ operator returns the top $c$ items in the item set $A$ in the descending order of $f(u, v)$. 

Denote $u \in \mathrm{R}^d$ as the user (or query) representation, and $v \in \mathcal{I}$ as the item representation, where $\mathcal{I}$ is the whole candidate set. Once an item $v$ is exposed to a user $u$, we observe corresponding Bernoulli feedback $Y(u, v) \in \{0, 1\}$, which is sampled from the underlying interest $r(u, v) \in [0, 1]$, thus $\mathbf{E}[Y(u, v)] = r(u, v)$.

Under the above four-stage system, given a user $u$, denote the corresponding candidate set and output item set of stage $i$ as $\mathcal{I}^{i-1}(u)$ and $\mathcal{I}^i(u)$ respectively, where $i \in [4]:=\{1,2,3,4\}$ and $\mathcal{I}^0 \equiv \mathcal{I}$, $|\mathcal{I}^i(u)| = c_i$. 
%Note both $\mathcal{I}^i(u)$ and $c_i$ may be different in different requests. 
For the clarity of our notations, we will use $\mathcal{I}^i$ instead of $\mathcal{I}^i(u)$ when there is no ambiguity. Given candidate set $\mathcal{I}^{i-1}$ at stage $i$, suppose item $v \in \mathcal{I}^{i-1}$ is selected into the output item set $\mathcal{I}^{i}$ with probability $p^{i}(u, v| \mathcal{I}^{i-1})$, and denote $O^i(u, v|\mathcal{I}^{i-1})$ as the corresponding observed Bernoulli random variable. Thus $\mathbf{E}[O^i(u, v|\mathcal{I}^{i-1})] = p^{i}(u, v| \mathcal{I}^{i-1})$. Here, we want to emphasize that $p^{i}(u, v| \mathcal{I}^{i-1})$ depends on $\mathcal{I}^{i-1}$, which means the probability of the same item that enters the next stage is also influenced by other candidate items in the same stage and reveals the combinatorial complexity of real-world systems. 
% Define $p^{i,j}(u, v):= p^i(u, v)p^{i+1}(u, v) \cdots p^j(u, v)$ and $O^{i,j}(u, v):= O^i(u, v)O^{i+1}(u, v) \cdots O^j(u, v)$, where $i < j$ and $i, j \in \{1, 2, 3,4\}$. Therefore whether an item $v$ is exposed to a user $u$ is observed by random variable $O^{1,4}(u, v)$ which obeys Bernoulli distribution with probability $p^{1, 4}(u, v)$.

%% file: gprp.tex
\section{Generalized Probability Ranking Principle}
\label{sec:gprp}
Before we introduce Generalized Probability Ranking Principle (GPRP), we recap the Probability Ranking Principle (PRP) in classical information retrieval \cite{robertson1977probability, wechsler2000probability} first:

\textit{Probability Ranking Principle (PRP): If an Information Retrieval system's response to each request is a ranking of items in the collections in order of decreasing probability of usefulness to the user who submitted the request, then the overall effectiveness of the system to its users will be the best that is obtainable on the basis of the data.}

The principle is intuitive and proved from the view of traditional measure of effectiveness and decision theory under certain assumptions  \cite{robertson1977probability}. Suppose we know the underlying interest function $r(u, v)$ in advance for every request and the system has only one stage (using the same notation as previous section that the system needs to return $c_4$ items to each request), according to PRP, then $c_4$ items returned by the system should be:
\begin{equation}
    \argmax_{\{v_k \in \mathcal{I} |k \in [c_4]\}} \sum_{k=1}^{c_4} r(u, v_{k}) \label{eq: one-stage prp}
\end{equation}

Suppose we are in the 4-stage system described before, and we need improve a particular stage $i \in \{1,2,3,4\}$ but have no control on other stages. If we still use PRP in stage $i$, we should output:
\begin{equation}
   \mathcal{I}^{i, PRP} := \argmax_{\mathcal{I}^{i} \subset \mathcal{I}^{i-1}} \sum_{v \in \mathcal{I}^{i}} r(u, v) \label{eq: multi-stage prp}
\end{equation}

Hence, nearly all previous studies in information retrieval systems focus on how to learn the underlying interest function $r(u, v)$ from logged exposure data, for either retrieval stage \cite{youtubeDNN,MIND_model,comirec,SASREC,TDM} or ranking stage \cite{youtubeDNN,liu2009learning}. Furthermore, since logged exposure data is affected by selection bias of the system which may lead to biased interest in learned models, some work further studies how to eliminate selection bias of the system to learn users' unbiased interests \cite{chen2019top,zhou2021contrastive,joachims2017unbiased,hu2019unbiased,saito2020unbiased}.

However, even though we could learn $r(u, v)$ perfectly, in a multi-stage system, the item set $\mathcal{I}^i$ ($i < 4$) cannot be exposed to users directly and it needs to be further filtrated by subsequent stages. In other words, even if we could find some items of high relevance to a user, once these items are filtered by subsequent stages, it would not bring any benefit since users can not see them. Thus, what we really care about in stage $i$ is the usefulness of final returned items after all subsequent stages, which means we should take both the preference of subsequent stages and the true interest of users into consideration at the same time. Hence we propose the following Generalized Probability Ranking Principle:

\textit{Generalized Probability Ranking Principle (GPRP): In the stage $i$ (any $i \in \{1,2,3\}$) of a four-stage system, if the output of stage $i$ to each request is a set which considers both the preference of subsequent stages and usefulness to the target user, as stated in the following equation (\ref{eq: multi-stage ideal gprp}), then the overall effectiveness of that stage to its users will be the best that is obtainable on the basis of the data. }
\begin{equation}
   \mathcal{I}^{i, GPRP} := \argmax_{\mathcal{I}^{i} \subset \mathcal{I}^{i-1}} \mathbf{E}[\sum_{v \in \mathcal{I}^{4}(\mathcal{I}^{i})} r(u, v)] \label{eq: multi-stage ideal gprp} 
\end{equation}
% \vspace{-2mm}
where $\mathcal{I}^{4}(\mathcal{I}^{i})$ is a random set that represents the finally exposed item set $\mathcal{I}^{4}$ obtained from the candidate set $\mathcal{I}^{i}$ after stages $i+1, \dots, 4$, and the expectation is over the randomness of subsequent stages.

According to the definition of equation (\ref{eq: multi-stage ideal gprp}), it's easy to see the expected utility of PRP is always a lower bound of the expected utility of GPRP when deployed in online systems:
\begin{equation}
    \mathbf{E}[\sum_{v \in \mathcal{I}^{4}(\mathcal{I}^{i, PRP})} r(u, v)] \leqslant \mathbf{E}[\sum_{v \in \mathcal{I}^{4}(\mathcal{I}^{i, GPRP})} r(u, v)] \label{eq: prp vs gprp} 
\end{equation}

To certain extend, the core philosophy behind GPRP is on the opposite to unbiased learning discussed in Section \ref{sec:related}. Both of them consider the selection bias of the system. However, the goal of unbiased learning is to eliminate selection bias and learn the underlying interest function of users, i.e. $r(u, v)$, which is then used to make decisions. While in GPRP of a multi-stage system, though underlying interest function $r(u, v)$ is important, the selection bias of subsequent stages is also important in determining returned items, and we should consider both of them when making decisions, which will lead to better final performance according to inequality (\ref{eq: prp vs gprp}). 

However, it is hard to formulate random set $\mathcal{I}^{4}(\mathcal{I}^{i})$ given the fact that subsequent stages after stage $i$ may contain both black-box algorithmic models and manual filtering strategies in practical systems, let alone solving equation (\ref{eq: multi-stage ideal gprp}). Hence, we should consider a simplified approximation of equation (\ref{eq: multi-stage ideal gprp}) which allows efficient algorithms. We find that PRP can be viewed as a max-min or conservative approximation of equation (\ref{eq: multi-stage ideal gprp}) as stated in the following proposition, which suggests we can completely ignore the complex process of subsequent stages and optimize its minimal utility in the worst case scenario:
\begin{proposition}
    $\mathcal{I}^{i, PRP}$ defined in equation (\ref{eq: multi-stage prp}) is also the solution to the following optimization problem:
    \begin{equation}
        \argmax_{\mathcal{I}^{i} \subset \mathcal{I}^{i-1}} \min_{\{v_k \in \mathcal{I}^{i} |k \in [c_4]\}} \sum_{k=1}^{c_4} r(u, v_k) \label{eq: max min prp} 
    \end{equation}
    \label{prop: max min representation}
\end{proposition}
\vspace{-5mm}
Because of space limitations, all the proof details are presented in the appendix.

We are interested in studying the gap between the final utility from PRP and GPRP, which is defined as
\begin{equation}
    \mathbf{E}[\sum_{v \in \mathcal{I}^{4}(\mathcal{I}^{i, GPRP})} r(u, v)] - \mathbf{E}[\sum_{v \in \mathcal{I}^{4}(\mathcal{I}^{i, PRP})} r(u, v)] \label{eq: gap}
\end{equation}

Apparently, when the system has only one stage or in the last stage of a multi-stage system, there are no subsequent stages, and the returned item set can be exposed to uses, hence GPRP degrades to PRP and the performance gap (\ref{eq: gap}) is 0. But we find the gap can be large in a non-final stage of a multi-stage system in the worst case without any assumption, as stated in the following proposition:
\begin{proposition}
    The performance gap (\ref{eq: gap}) between PRP and GPRP can be $c_4$ in the worst case. \label{prop: worst case result}
\end{proposition}

Note the upper bound of equation (\ref{eq: gap}) is exactly $c_4$, which means PRP is not a good approximation of GPRP in the worst case of a multi-stage system and thus we have to find some other proxy for the objective function in equation (\ref{eq: multi-stage ideal gprp}).

Since $\mathcal{I}^{4}(\mathcal{I}^{i})$ is a random set depending on the input candidate set, and the probability of each item $p^{i}(u, v| \mathcal{I}^{i-1})$ being selected into the next stage $i+1$ is also influenced by other candidates, which causes the computational complexity of solving equation (\ref{eq: multi-stage ideal gprp}) exactly. We then make the following assumption to allow efficient approximate solution to equation (\ref{eq: multi-stage ideal gprp}):
\begin{assumption}
    Assume the selection probability $p^{i}(u, v| \mathcal{I}^{i-1})$ depends only on the user and item itself, i.e. $p^{i}(u, v| \mathcal{I}^{i-1}) = p^{i}(u, v)$.
    \label{assumpution one}
\end{assumption}

This is a rather strong assumption in practical systems at first glance, since general ranking stages need to compare all input candidates together, hence we cannot ignore the influence of other candidate items in each request. We will discuss the validity of this assumption in detail in Section \ref{sec:assum verify}.

Under Assumption \ref{assumpution one}, we can rewrite the objective function (\ref{eq: multi-stage ideal gprp}) of GPRP into the following:
\begin{equation}
   \bar{\mathcal{I}}^{i, GPRP} := \argmax_{\mathcal{I}^{i} \subset \mathcal{I}^{i-1}} \sum_{v \in \mathcal{I}^{i}} p^{i+1, 4}(u, v)r(u, v) \label{eq: multi-stage ideal point-wise gprp} 
\end{equation}
where $p^{i,j}(u, v):= p^i(u, v)p^{i+1}(u, v) \cdots p^j(u, v)$. Here we use $\bar{\mathcal{I}}^{i, GPRP}$ to distinguish $\mathcal{I}^{i, GPRP}$ in equation (\ref{eq: multi-stage ideal gprp}). When Assumption \ref{assumpution one} holds, they are the same, otherwise they are different.

Compared with the original objective function (\ref{eq: multi-stage ideal gprp}), the new objective function looks more like the objective function (\ref{eq: multi-stage prp}) of PRP, which gets rid of the annoying combinatorial form of subsequent stages and allows efficient learning like many previous algorithms under PRP. However, different from the objective function (\ref{eq: multi-stage prp}) of PRP which only considers the interest of users, the new objective function (\ref{eq: multi-stage ideal point-wise gprp}) still considers both the selection bias of subsequent stages (i.e. $p^{i+1, 4}(u, v)$ term) and the interest of users (i.e. $r(u, v)$ term), which inherits the merit of GPRP. 

Without Assumption (\ref{assumpution one}), though it is hard to compare $\bar{\mathcal{I}}^{i, GPRP}$ in (\ref{eq: multi-stage ideal point-wise gprp}) and $\mathcal{I}^{i, PRP}$ in (\ref{eq: multi-stage prp}) in general, we give two observations in some special cases of the system, which may help us understand the relation between them.

\begin{obser}
    If the selection bias of subsequent stages $p^{i+1, 4}(u, v)$ is independent of items, then $\bar{\mathcal{I}}^{i, GPRP} = \mathcal{I}^{i, PRP}$. \label{obs: one}
\end{obser}

The observation is obvious. For example, when subsequent stages' output candidate items to the next stage are completely random, then the optimal candidate set should be chosen according to the underlying interest of users. However, the completely random strategy is very dangerous in an online environment which can hurt users' experience on the platform, and the strategy in each stage is often some learned models from logged data.

\begin{obser}
    If the selection bias of subsequent stages $p^{i+1, 4}(u, v)$ is monotonically increasing with respect to the underlying interest $r(u, v)$ , then $\bar{\mathcal{I}}^{i, GPRP} = \mathcal{I}^{i, PRP}$. \label{obs: two}
\end{obser}

Above observation is also straightforward, since $p^{i+1, 4}(u, v)$ is monotonically increasing with $r(u, v)$, which implies the ranking of items in terms of $p^{i+1, 4}(u, v)r(u, v)$ is the same as the ranking by $r(u, v)$, thus $\bar{\mathcal{I}}^{i, GPRP}$ equals $\mathcal{I}^{i, PRP}$. However, the assumption in observation \ref{obs: two} is too strict, as it requires the preference of each stage to exactly coincide with the interest of users in every request, which is impractical considering the existence of learning error of any algorithm in any stage.

Now, given the difference between objective functions (\ref{eq: multi-stage ideal point-wise gprp}) and (\ref{eq: multi-stage prp}), the key question is then how to solve equation (\ref{eq: multi-stage ideal point-wise gprp}) efficiently. A straightforward method is to learn $p^{i+1, 4}(u, v)$ and $r(u, v)$ separately, where each term in $p^{i+1, 4}(u, v)$ can be learned by any supervised algorithm on corresponding data and $r(u, v)$ term can be learned by any existing algorithm especially unbiased learning algorithms \cite{chen2019top,zhou2021contrastive,joachims2017unbiased,hu2019unbiased,saito2020unbiased}. Having learned $p^{i+1, 4}(u, v)$ and $r(u, v)$, we can sort items in $\mathcal{I}^{i-1}$ according to $p^{i+1, 4}(u, v)r(u, v)$ and output the top $c_i$ items as $\bar{\mathcal{I}}^{i, GPRP}$. However, this method is also inefficient and may be ineffective, as we need to learn several models which causes additional computational burden and accumulated learning error. Besides, in the retrieval stage, usually we use Approximate Nearest Neighbor (ANN) for fast inference, which does not support the operation in equation (\ref{eq: multi-stage ideal point-wise gprp}). To address the above issue, we propose a unified algorithmic framework named Full Stage Learning to Rank.

%% file: method.tex
\section{Full Stage Learning to Rank}
\label{sec:method}
According to objective function (\ref{eq: multi-stage ideal point-wise gprp}), we need to sort items according to $p^{i+1, 4}(u, v)r(u, v)$. Instead of learning each term independently, we directly learn the product $w^{i+1, 4}(u, v) := p^{i+1, 4}(u, v)r(u, v)$ to avoid issues mentioned in previous section. Apparently $O^{i+1, 4}(u, v)Y(u, v)$ is a realization of $w^{i+1, 4}(u, v)$ which can be observed in practical systems. As $\mathbf{E}[O^{i+1, 4}(u, v)Y(u, v)] = w^{i+1, 4}(u, v)$, we obtain our first efficient algorithm towards learning objective function (\ref{eq: multi-stage ideal point-wise gprp}) of GPRP: collecting data after stages $i-1$ with label 0, except the exposed and clicked data with label 1, then learning a supervised model with such labeled data which is then used online at stage $i$.

Though this algorithm is simple and easy to implement, it treats all non-exposed items equally, which may omit the abundant information of the data. Intuitively, items which enter stage $i+1$ may be better than items filtrated by stage $i$, thus we hope to make full use of the information of collected data to learn the selection bias and user's interest better. 

Before introducing our new algorithmic framework, we make another important assumption about the multi-stage system:

\begin{assumption}
    For two observed items $v_i, v_{i+1}$ in two subsequent stages $i$ and $i+1$ (i.e. $v_i \in \mathcal{I}^{i}, v_i \not\in \mathcal{I}^{i+1}, v_{i+1} \in \mathcal{I}^{i+1}$), where $i \in [3]$, the ratio between user's interests of these two items is bounded in a constant interval, which is less than the ratio of selecting probability between them, that is
    \begin{equation}
        \frac{1}{a} \leqslant \frac{r(u, v_{i+1})}{r(u, v_i)} \leqslant a \leqslant \frac{p^{i+1:4}(u, v_{i+1})}{p^{i+1:4}(u, v_i)}
        \label{eq: stage probability ineq}
    \end{equation}
    where $a \geqslant 1$ is a constant. 
    \label{assum: two}
\end{assumption}

The intuition behind Assumption \ref{assum: two} is that each stage of the system can be seen as a cluster of items depending on the degree of user's interest in some sense. Apparently, for an information retrieval system with good performance, user's interest for items in higher stages may be stronger than interest for items in lower stages with high probability. What's more, the interest of users with respect to two items in subsequent stages are close to each other. Note this condition does not require the selection bias of the system is exactly the same with user's underlying interest, and allows the learning error of system to some degree, which is much weaker than the monotone assumption of multi-stage systems used in Observation \ref{obs: two} and in line with practical situation. See more discussion about this assumption in the next section, where we show it is relatively reasonable in practical multi-stage systems.

For $t$-th request of the system, we collect a series of data at all stages $\{(u^t, v^t_k, S^t_k, \bar{Y}^t_k)| k \in [M]\}$, where $M$ is the number of collected data in each request and $S_k^t \in [4]$ is an observed random variable representing the stage of item $v_k^t$. $\bar{Y}_k \in \{0, 1, NA\}$ here represents the feedback of $(u, v_k)$ pair, where $\bar{Y}_k = Y_k$ when $S_k = 4$, otherwise $\bar{Y}_k = NA$, since we can only observe the true feedback $Y_k(u, v_k)$ when the item is exposed to the user.

With these data collected from full stages of different requests, the main technique of our method is to relabel each user-item pair $(u, v, S, \bar{Y})$ in collected data set by the following rule:
\begin{align}
    L(u, v) = \left\{\begin{array}{ll}{z_0} &
    {\text{if } S(u, v) = 0 } \\
    {z_i} & {\text{if } S(u, v) = i, \text{for } i < 4} \\
    {z_4} & {\text{if } S(u, v) = 4, \text{and } Y(u, v) = 0} \\
    {z_5} & {\text{if } S(u, v) = 4, \text{and } Y(u, v) = 1}
    \end{array}\right.
    \label{eq: relabel}
\end{align}
where $z_0 \leq z_1 \leq \dots \leq z_5$ are six non-negative numbers.

The core idea behind this relabeling technique is intuitive, which distinguish the difference among non-exposed items and makes full use of them. Items which are returned to users and receive positive feedback apparently are the most important ones, since they have passes the examination of selection bias of subsequent stages and enjoys users interest. For those items that have been retrieved but not shown to users, entering into next stage of the system is more difficult compared with being liked by users. Therefore, if an item enters into a higher stage of the system, it is more important compared with items in lower stages. In fact, under Assumption \ref{assumpution one} and \ref{assum: two}, one can prove above relabeling technique coincides with GPRP as stated in the following theorem, which give us a theoretical support of our simple relabeling technique.
% \vspace{-2mm}
\begin{theorem}
    Suppose Assumption \ref{assumpution one} and \ref{assum: two} hold in multi-stage systems, for any collected data $\{(u, v_k, S_k, \bar{Y}_k)| k \in [M]\}$, after relabeling via equation (\ref{eq: relabel}), we obtain new labels $L(u, v_k)$ for each data respectively, then a ranking of items in decreasing order of $L(u, v_k)$ implies the ranking by $w^{i+1, 4}(u, v_k)$ for items in different stages. 

    % \begin{equation}
    %     \mathcal{I}^{i, GPRP} = \argmax_{\mathcal{I}^{i} \subset \mathcal{I}^{i-1}} \sum_{v \in \mathcal{I}^{i}} \mathbf{E}[L(u, v)] \label{eq: multi-stage practical point-wise gprp} 
    % \end{equation}
    \label{thr: main}
\end{theorem}
% \vspace{-2mm}
According to Theorem \ref{thr: main},  now we can use the new label $L(u, v)$ as an approximate substitution of $w^{i+1, 4}(u, v)$, and this new relabeling also allows efficient learning. Once we can estimate $L(u, v)$ or has the ability of ranking items which is the same as ranking by $L(u, v)$, then it nearly matches GPRP, since $\mathcal{I}^{i, GPRP}$ is obtained by ranking $w^{i+1, 4}(u, v)$ in decreasing order according to equation (\ref{eq: multi-stage ideal point-wise gprp}) under Assumption \ref{assumpution one}. Thus, we can use any supervised or general ranking algorithm to reach an efficiently learned model that coincides with GPRP. In detail, having obtained the collected relabeled data $\{\{(u^t, v^t_k, L^t_k)| k \in [M]\}| t \in [N]\}$, where $N$ is the total number of collected requests, we can use any supervised learning or Learning to Rank (LTR) algorithm to learn a model with the goal of correctly ranking items $\{(u^t, v^t_k, L^t_k)| k \in [M]\}$ in each request, for example Lambda Rank \cite{burges2006learning,burges2010ranknet}. 

At stage $i$, since we only care about the selection bias of stages after $i$ and users' underlying interest, the collected data at stage $i$ and before stage $i$ seems useless. However, from the view of consistency between training and serving, it will be better to use them duration training. For example when serving at retrieval stage, we use the learned model to predict scores for all items. Therefore, it will help a lot to add some random samples from the candidate pool during training. The same reasoning applies at other stages. What's more, we can also add the collected data at any other previous stage duration training, which can be regarded as an auxiliary training task and may help learn the model better. Note using data in previous stages doesn't influence the near consistency between our algorithm and GPRP.

The final algorithmic framework including training and inference is given in Algorithm \ref{Algo: FS-LTR}. Since we need collect data from full stages of a system and use LTR algorithm as our backbone, we name our algorithmic framework as Full Stage Learning to Rank, FS-LTR in short. When $S_k^t=0$ in Algorithm \ref{Algo: FS-LTR}, it means the item $v_k^t$ is randomly sampled from $\mathcal{I}$. 

\begin{algorithm}[t]
\caption{Full Stage Learning to Rank}
\label{Algo: FS-LTR}
\SetKwFunction{}
\KwFunction{\textbf{Training:}} \\
\KwIn{
Initialized model parameter $\theta$, manually defined labels $\{z_i | i \in [0, 5]\}$ and original data-set  $\{\{(u^t, v^t_k, S^t_k, \bar{Y}_k^t)| k \in [M], S^t_k \geq j\}| t \in [N]\}$, where $j \in [0, i+1]$ is a hyper-parameter.
% , where $N$ is the number of requests and $M$ is the number of collected data in each request. 
}
\KwOut{Model $\hat{l}(u, v | \theta)$}

\SetAlgoLined
\BlankLine
\hspace{2em}\textbf{While $\theta$ not converged} \textbf{do}\\
\hspace{3em} sample a batch of requests from $[N]$\\
\hspace{3em} relabel each user-item pair by the equation (\ref{eq: relabel})\\
\hspace{3em} update model parameters by chosen LTR algorithm \\
\hspace{2em}\textbf{return} $\theta$ \\
\SetKwFunction{}
\KwFunction{\textbf{Serving in stage $i$:}} \\
\KwIn{
Candidate set $\mathcal{I}^{i-1}$ in a request of user $u$, learned model parameters $\theta$
}
\KwOut{ $\mathcal{I}^{i}$}

\SetAlgoLined
\BlankLine
\hspace{2em} $\mathcal{I}^{i} := \mathbf{Topk}(\mathcal{I}^{i-1}, \hat{l}(u, v | \theta), c_i)$\\
\hspace{2em}\textbf{return} $\mathcal{I}^{i}$
\vspace{-1mm}
\end{algorithm}

As discussed above, by choosing different $j$ and appropriate models, Algorithm \ref{Algo: FS-LTR} can be used either in retrieval, pre-ranking, ranking or re-ranking. For example, in retrieval stage, we can choose $j=0$ which includes random negative items from the whole candidate set and dual-encoder models which represents users and items as vectors and supports ANN for fast inference. Similarly in pre-ranking/ranking stage, we can choose suitable $j$ and dual-encoder model or complex dnn models. In re-ranking stage, Algorithm \ref{Algo: FS-LTR} degrades to backbone LTR algorithm if only using exposed data, and becomes a new LTR algorithm with auxiliary tasks which uses data in previous stages.

%% file: assum_verify.tex
\section{Discussion about Assumptions}
\label{sec:assum verify}
As mentioned in Section \ref{sec:gprp}, Assumption \ref{assumpution one} is too strong to be true in real complex multi-stage systems, but it is acceptable if this assumption could be satisfied in some degrees, and we explain its reasonableness from four viewpoints:
% \vspace{-0.8mm}
% \begin{itemize}

[1] The simplification from combinatorial system into point-wise system is quite common in similar situation, like in re-ranking \cite{ie2019slateq}.
    % \item[2] Most of previous work about unbiased learning \cite{chen2019top} are based on estimating point-wise exposure probability to selection exposure bias of the system and then learn the underlying interests, which also rely on Assumption \ref{assumpution one} implicitly.  

[2] Though the probability $p^{i}(u, v| \mathcal{I}^{i-1})$ varies with input candidate set $\mathcal{I}^{i-1}$, we may consider the general performance of each user-item pair, i.e. $p^{i}(u, v) = \mathbf{E}[p^{i}(u, v| \mathcal{I}^{i-1})]$ as an approximation to $p^{i}(u, v| \mathcal{I}^{i-1})$, where the expectation is over some distribution of $\mathcal{I}^{i-1}$ (for example the distribution of $\mathcal{I}^{i-1}$ of current system).

[3] We have learned several binary classifiers based on samples collected from each stage in our online system to classify whether the input candidate item in each stage could enter the next stage. Then we test these classifiers on the evaluation dataset. We find all metrics, including AUC/UAUC/WUAUC, are above 0.8 (some even above 0.9), which strongly suggests we can learn the selection probability of each stage in a point-wise manner only based on the user and item.

[4] We only use Assumption \ref{assumpution one} to induce efficient and practical algorithm. In experimental section, we find our algorithm still works when this assumption does not hold either in simulated experiments or online A/B testing.
% \end{itemize}

Assumption \ref{assum: two} is also critical for FS-LTR. However, it is impossible to collect real $r(u, v)$ and $p^i(u, v)$ in real systems, as we cannot present the same item twice for a user in short-video platform which implies we can only observe one realization $Y(u, v)$ of underlying Bernoulli distribution with probability $r(u, v)$. What's more, items at non-final stages of the system are not exposed to users, so we cannot even observe their realizations of users' underlying interest. Besides, practical system is very complex, there does not exist any true $p^i(u, v)$ in real situation. 

One possible approach to solving above difficulty is to learn approximate models about $r(u, v)$ and $\{p^i(u, v)|i \in [4]\}$ respectively. However, this approach highly depends on the learning performance which is also hard to have some guarantee, because of the discrepancy between learning space and inference space. Therefore, we adopt an approximate verification approach. 

In detail, we collect items at different stages in previous request (there are 5 stages in our system), and then force these items to be exposed to corresponding users directly, which don't need to enter the multi-stage system to avoid its selection bias. Now we can receive ground-truth feedback of items at different stages. Table \ref{tab: posterior ctr} shows the average performance of XTR (mainly Click/Like/Follow Through Rate) at different stages of such collected data. We can see average posterior XTR is very close between consecutive stages, and the ratio between them is also in a small interval $[\frac{1}{2}, 2]$. Thus we know users' interest for all items which have entered the system are close at least on average, which implies the inequality $ \frac{1}{a} \leqslant \frac{r(u, v_{i+1})}{r(u, v_i)} \leqslant a$ of equation (\ref{eq: stage probability ineq}) may be true in real-world and $a$ may be 2 in our case. Besides, the inequality about term $\frac{p^{i+1:4}(u, v_{i+1})}{p^{i+1:4}(u, v_i)}$ in equation (\ref{eq: stage probability ineq}) means an item that could be selected to the next stage is relatively more difficult than user whether likes it compared with other candidate items in subsequent stages. In practical system where $c_1/c_2/c_3/c_4/c_5 = 6000/3000/500/120/10$, the selecting probability from stage 2 to 3 is roughly $\frac{500}{3000}=\frac{1}{6}$ on average, which means the ratio of selecting probabilities between positive (i.e. select by stage 2 into stage 3) and negative item (i.e. filtrated by stage 2) in this stage is roughly 6 on average, hence the ratio of exposure probability between these two items after subsequent stages could be even larger than 6, which may be greater than $\frac{r(u, v_{i+1})}{r(u, v_i)}$ with high probability. Therefore, Assumption \ref{assum: two} may be mild enough to be satisfied in a real environment, at least in a short-video platform. 

\begin{table}[t]
    \centering
    \caption{Posterior XTR of Items at Different Stages.}
    \begin{tabular}{c|ccccc}
         \toprule
         & Stage 1 & Stage 2 & Stage 3 & Stage 4 & Stage 5 \\
         \midrule
         \textit{Posterior CTR} & 0.29 & 0.33 & 0.34 & 0.49 & 0.56 \\
         \midrule
         \textit{Posterior LTR} & 0.009 & 0.010 & 0.015 & 0.022 & 0.026 \\
         \midrule
         \textit{Posterior FTR} & 0.0018 & 0.0022 & 0.0029 & 0.0038 & 0.0046 \\
         \bottomrule
    \end{tabular}
    \label{tab: posterior ctr}
\end{table}

%% file: practice.tex
\section{Practical Implementation}
\label{sec:practice}

In this section, we present our implementation details of FS-LTR in real-world online multi-stage recommendation systems, including data collection, training, and serving.

\noindent \textbf{Data Collection:} Collecting data in full stages of a recommendation system online is a very challenging engineering task, since the number of these data far exceeds the magnitude of the exposure data in each request. For example, it could be 10000 versus 6 in real system, hence it is impractical to collect all of them considering the cost of storage and communication bandwidth. In practice, for each requst, we randomly sample 40 negative items in retrieval and pre-ranking stage respectively, and record all items (around 400) in ranking stage with corresponding concrete ranking information.

\noindent \textbf{Training:} We use lambda rank to train our model in retrieval stage with $z_i:=i$. To reduce training cost, we only use a subset of collected data in each request. The higher stage the data belongs to, the more we use it, since data in higher stages is more important and hard to be learned. We also substitute the cross-entropy loss in lambda rank with margin loss, as well as trying different labels (for example, 0, 1, 2, 3, 4, 6) to enhance the learning of hard samples. Though having some improvements in offline evaluation, it doesn't lead to online gain. What's more, to speed up training, one could use pairwise LTR instead of lambda rank, which avoids expensive sort operation during training but with only mild damage in online performance. Finally, it is also possible to use softmax loss with soft labels $l(u, v)$ to further speed up training.

\noindent \textbf{Serving:} At retrieval stage, to enhance the diversity of returned results in each request, we add some noise to the top representation vector of users, which also fulfills some mild exploration. The same trick could be used in other stages too. 

%% file: experiment.tex
\section{Experiments}
\label{sec:experiments}
In this section, we conduct both offline experiments and online A/B testing to answer the following research questions:
\begin{itemize}
    \item[\textbf{RQ1}] How can we simulate a multi-stage system with an offline dataset for evaluating our proposed method? (Sec \ref{sec:6.1})
    \item[\textbf{RQ2}] What is the effectiveness of FS-LTR on the offline simulated multi-stage recommendation system? (Sec \ref{sec:6.2})
    \item[\textbf{RQ3}] How does each component or hyperparameter in Full Stage Learning to Rank affect the performance? (Sec \ref{sec:6.3})
    \item[\textbf{RQ4}] How does FS-LTR perform in online environment? (Sec \ref{sec:6.4})
\end{itemize}

\subsection{Offline Simulation of Multi-Stage Pipeline}
\label{sec:6.1}
In this section, we introduce our offline simulation from two perspectives: dataset, multi-stage simulation.

\subsubsection{Dataset} Different stages of a multi-stage recommendation have a huge magnitude difference in the number of candidates, thereby the first stage needs an enormous amount of candidates. However, most of the datasets in recommendation systems are highly sparse, which leads to a limited number of available reasonable candidates. Even if we can train a model to capture the latent preference of the user, without the ground truth of user-item interaction, we still cannot determine whether the user prefers the item. The original sparse ground truth may be more sparse after passing the multi-stage pipeline.

\begin{table}
    \centering
    \caption{Statistics of KuaiRec dataset.}
    % \vspace*{-\baselineskip}
    \begin{tabular}{c|cccc}
         \toprule
         % \hline
         & \#User & \#Item & \#Interaction & Density\\
         \midrule
         \textit{small matrix} & 1,411 & 3,327 & 4,676,570 & 99.6\% \\
         \textit{big matrix} & 7,176 & 10,728 & 12,530,806 & 16.3\% \\
         % \hline
         \bottomrule
    \end{tabular}
    \label{tab:stat}
\end{table}
To build a convincing simulation on the multi-stage recommendation pipeline, we utilize the fully-observed dataset \textbf{KuaiRec}\cite{gao2022kuairec} as our base dataset. KuaiRec is a real-world dataset collected from the recommendation logs of a video-sharing platform. "Fully Observed" means that there are almost no missing values in the user-item interaction matrix, allowing an enormous amount of available candidates with ground truth for simulating the multi-stage recommendation pipeline. There are two user-item interaction matrices in KuaiRec, named \textit{small matrix} and \textit{big matrix}. The statistics of the two matrices are listed in Table \ref{tab:stat}. Both of these two matrix are more dense than most of the recommendation datasets. Moreover, rich side features are provided for each user and item in the KuaiRec dataset, enabling training a good ranking model for simulating the multi-stage pipeline. All of the user and item in \textit{small matrix} also occur in the \textit{big matrix}, but interactions in \textit{small matrix} and \textit{big matrix} are excluded from each other.

\subsubsection{Multi-Stage Simulation} 
We consider building a multi-stage pipeline containing three parts: retrieval, prerank, and rank. Each item must pass all these three stages to be exposed to the user. In each stage, a learned model will score each candidate item, and items with relatively high scores can be passed to the next stage. We assume that
%\begin{itemize}

     $\bullet$ \textbf{Retrieval candidate pool} contains all of the items in \textit{big matrix}( \textasciitilde11000).
    
     $\bullet$ \textbf{Prerank candidate pool} of each user contains all of the items exposed to the user in both \textit{big matrix} and \textit{small matrix}(\textasciitilde3500).

    % ~\\
%\end{itemize}
Since we hope the learned models have a relatively strong ability to predict users' interests, we use the \textit{big matrix} with user features and item features to train the prerank / rank model. We select the two-tower DNN for the prerank model and common DNN for the rank model. Details can be viewed in Appendix \ref{appendix: pipeline}. All of the prerank candidates can be scored with the prerank model and the rank model, enabling us to simulate the multi-stage pipeline. For the convenience of later training and evaluation, we model the whole multi-stage pipeline as a static \textbf{request}, which approximately simulates our online practical system. See \ref{appendix: pipeline} for more details about data generation of a static request. 

Multiple request procedures can be repeatedly simulated to generate different multi-stage recommendation training samples. For evaluation, requests for validation or tests are separated in order to avoid label leakage. Within this simulated environment, see Appendix \ref{appendix: data preparation} about concrete data preparation for training and evaluation. 

\subsection{Performance Comparison}
\label{sec:6.2}
\subsubsection{Implementation Details}
We utilize the \textbf{MF}\cite{koren2009matrix} as the model structure for all methods. We implement all of the offline experiments with Pytorch 1.13 in Python 3.8. We set the same value for basic hyperparameters. The size of the embedding is set to 64, the batch size is set to 128. L2 normalization is applied to the embedding during training and inference. We use the Adam optimizer with a learning rate of 0.001 for training for all methods. 

\subsubsection{Baseline Methods} 
We select multiple widely used optimization objectives as baseline methods for our experiments:
\begin{itemize}
    \item \textbf{BPR}\cite{rendle2012bpr} (Beyesian Personalized Ranking). An objective models the posterior preference probability of a user-item pair based on a single sampling negative user-item pair. The default number of negative samples is 5.
    \item \textbf{SSM}\cite{covington2016deep,jean2014using} (Sampled-SoftMax). An objective models the preference probability of user-item pairs based on multiple sampling negative negative user-item pairs. The default number of negative samples is 5.
    \item \textbf{BCE} (Binary Cross Entropy). An objective directly optimizes the preference probability of the user-item pair.
    \item \textbf{RF} (RankFlow\cite{qin2022rankflow}). A joint training framework for optimizing multi-stage cascade ranking system \footnote{We have removed the mean squared error (MSE) loss term from the RankFlow objective function, as we found that it only introduced worse performance in our setting.}.
\end{itemize}
BPR and SSM are common baseline methods for retrieval, and negative user-item pairs for them are sampled from the global candidate pool. BCE is adopted as the baseline method for our offline experiment on prerank stage.

\subsubsection{Evaluation Metrics}
We adopt four widely used metrics for evaluation of our methods, i.e., \textbf{R} (Recall), \textbf{P} (Precision), \textbf{HR} (HitRate), and \textbf{NDCG} (Normalized Discounted Cumulative Gain). The metrics are computed on the top 20/50 matched items. As mentioned above, the different stage has a different candidate pool for evaluation. The prerank stage has a relatively smaller candidate pool than retrieval. However, metrics only have a very limited difference between the retrieval and prerank stage for our imperfect simulation on a multi-stage recommendation system.

\subsubsection{Overall Performance} 
Performance comparison is listed in Table \ref{tab:perf}. For baseline methods and our FS-LTR methods in retrieval, the negative sample size for each training instance is set to 5 to eliminate the effect of the negative sample size. Two variants of our FS-LTR, FS-RN(RankNet) and FS-LR(LambdaRank) outperform significantly to the baseline methods in the retrieval stage. Our FS-LR method also shows an advantage over the commonly used BCE in the prerank stage.
% \noindent\makebox[\textwidth][s]{
\begin{table*}
    \centering
    \caption{Performance comparison on our simulated multi-stage pipeline. Bold numbers represent the best results. All the numbers in the table are percentage numbers with `\%' omitted.  All experiments are repeated 5 times to calculate the mean and standard deviation. We conduct an unpaired t-test of Full Stage Learning to Rank and the best baseline and the improvement of Full Stage Learning to Rank is significant with $p \leq 0.05$ for all values with `\textbf{*}'.}
    \resizebox{\textwidth}{15mm}{
    % \begin{tabularx}{\textwidth}{c|c|cccc|cccc}
    \begin{tabular}{c|c|cccc|cccc}
    \toprule
    \hline
    Stage & Method & HR@20 & NDCG@20 & R@20 & P@20 & HR@50 & NDCG@50 & R@50 & P@50 \\ \midrule
        \multirow{4}{*}{Retr} & BPR & 15.77($\pm$0.97) & 5.50($\pm$0.53) & 0.55($\pm$0.03) & 0.88($\pm$0.05) & 35.59($\pm$1.85) & 9.46($\pm$0.62) & 1.41($\pm$0.07)& 0.91($\pm$0.04) \\
        & SSM & 17.32($\pm$1.34) & 6.02($\pm$0.50) & 0.59($\pm$0.05) & 0.96($\pm$0.08) & 37.88($\pm$0.86) & 10.16($\pm$0.42) & 1.53($\pm$0.05) & 0.99($\pm$0.03) \\
        & RankFlow & 17.76($\pm$0.51) & 6.20($\pm$0.35) & 0.62($\pm$0.02) & 1.00($\pm$0.04) & 38.64($\pm$1.63) & 10.40($\pm$0.49) & 1.60($\pm$0.08) & 1.02($\pm$0.05) \\
        \cline{2-10}
        & FS-RN & 18.14($\pm$0.49) & 6.41($\pm$0.25) & 0.63($\pm$0.03) & 1.03($\pm$0.03) & 38.60($\pm$0.73) & 10.52($\pm$0.19) & 1.61($\pm$0.05) & 1.04($\pm$0.03) \\
        & FS-LR & \textbf{18.83($\pm$0.58)} & \textbf{6.69($\pm$0.12)}\textbf{*} & \textbf{0.66($\pm$0.02)} & \textbf{1.06($\pm$0.03)} & \textbf{40.75($\pm$0.33)}\textbf{*} & \textbf{11.07($\pm$0.08)}\textbf{*} & \textbf{1.72($\pm$0.02)}\textbf{*} & \textbf{1.10($\pm$0.01)}\textbf{*} \\\hline

        \multirow{2}{*}{PR} & BCE & 14.40($\pm$1.10) & 4.93 ($\pm$0.35) & 0.48($\pm$0.04) & 0.76($\pm$0.06) & 32.57($\pm$1.68) & 8.54($\pm$0.50) & 1.24 ($\pm$0.09) & 0.79($\pm$0.06) \\
        & RankFlow & 18.23($\pm$0.55) & 6.53($\pm$0.35) & 0.65($\pm$0.01) & 1.04($\pm$0.02) & 39.22($\pm$0.70) & 10.80($\pm$0.28) & 1.70($\pm$0.04) & 1.08($\pm$0.03) \\
        \cline{2-10}
        & FS-LR & \textbf{19.56($\pm$1.18)}\textbf{*} & \textbf{6.74($\pm$0.48)}\textbf{*} & \textbf{0.70($\pm$0.05)}\textbf{*} & \textbf{1.13($\pm$0.07)}\textbf{*} & \textbf{40.16($\pm$0.72)}\textbf{*} & \textbf{10.95($\pm$0.39)}\textbf{*} & \textbf{1.79($\pm$0.04)}\textbf{*} & \textbf{1.15($\pm$0.02)}\textbf{*} \\
    \bottomrule
    \end{tabular}}
    \label{tab:perf}
\end{table*}
\subsection{Ablation Study}
Now we conduct experiments to further understand the effect of multi-stage negative samples and negative sample size in FS-LTR.

\subsubsection{Effectiveness of multi-stage samples} 
The complete FS-LTR in retrieval requires user-item pairs from full stages in recommendation systems. We remove user-item pairs from different stages to verify the effectiveness of each stage sample. Results are shown in Table \ref{tab:diff_stage}. The re-labeled label is also adapted to the number of stages left in the training samples. Removing the stage-negative samples causes the most degradation in model performance, showing that negative samples in a multi-stage recommendation system play a key role in FS-LTR.

\subsubsection{The number of samples from different stages}
\label{sec:6.3}
We show the performance of FS-LR on the number of samples from different stages in Table \ref{tab:diff_num}. Results show that more negative sample sizes may have a positive effect on model performance. The ratio of different stage negative sample sizes is also critical for model performance. 

\begin{table}
    \centering
    \caption{Ablation study on different stage examples.}
    \begin{tabular}{c|cccc}
        \toprule\hline
         & HR@50 & NDCG@50 & R@50 & P@50 \\ \midrule
         w/o \textit{exposed neg} & 37.29 & 10.03 & 1.52 & 0.98\\
         w/o \textit{stage neg} & 35.52 & 9.72 & 1.44 & 0.93 \\
         w/o \textit{rank neg} & 39.40 & 10.74 & 1.63 & 1.04 \\
         w/o \textit{prerank neg} & 38.31 & 10.16 & 1.56 & 1.01 \\ \hline
         FS-LR & 40.75 & 11.07 & 1.72 & 1.10 \\ \hline
         \bottomrule
    \end{tabular}
    \label{tab:diff_stage}
\end{table}

\begin{table}
    \centering
    \caption{Study on the number of different stage examples. The `x,y,z' in the Negative Sampling column denotes the number of rank/prerank/global negative samples in one training instance respectively. Exposed negative sample size is set to 1 for all methods.}
    \begin{tabular}{c|cccc}
        \toprule\hline
         Negative Sampling & HR@50 & NDCG@50 & R@50 & P@50 \\ \midrule
         1,1,1 & 39.12 & 10.45 & 1.68 & 1.07 \\
         1,2,1 & 40.71 & 10.73 & \textbf{1.73} & \textbf{1.10} \\
         2,1,1 & 39.67 & 10.71 & 1.68 & 1.07 \\
         1,1,2 & \textbf{40.75} & \textbf{11.07} & 1.72 & 1.10 \\ 
         1,1,3 & 39.24 & 10.73 & 1.63 & 1.05 \\
         1,2,3 & 39.71 & 11.00 & 1.69 & 1.09 \\
         \bottomrule
    \end{tabular}
    \label{tab:diff_num}
\end{table}

\subsection{Online A/B Testing} 
\label{sec:6.4}
We used our Algorithm \ref{Algo: FS-LTR} at the retrieval stage on one of the largest short-video platforms with implementation details described in Section \ref{sec:practice}. The A/B test lasted for six days and had influenced over 20 million users in the experiment group, reaching a significant improvement compared with the base group which already had some strong baselines like TDM \cite{TDM}, Multi-Interest Retrieval \cite{zhang2023divide}, Comi-Rec Retrieval \cite{comirec} etc. As shown in Table \ref{tab:online}, our approach achieves significant gains in many engagement metrics which are statistically significant with under bootstrap test. What's more, our retrieval algorithm has the highest reveal ratio (16\%, where the second highest reveal ratio is 6\%) of any other retrieval algorithm online, which coincides with the GPRP.

\begin{table*}[h]
    \centering
    \caption{Online A/B testing results.}
    \begin{tabular}{c|ccccccc}
         \toprule
         % \hline
         App Usage Time Per User& Real-show & Click & Like & Follow & Forward & Comment & Watch Time \\
         \midrule
         +0.12\% & +0.50\% & +0.69\% & +0.40\% & 0.74\% & +0.94\% & 1.08\% & +0.18\% \\
         % \hline
         \bottomrule
    \end{tabular}
    \label{tab:online}
\end{table*}

% \begin{table}[]
%     \centering
%     \caption{Study on the number of different stage examples.'[x,y,z]' denotes the number of rank negative samples, prerank negative samples and global negative samples in one training instance, respectively.}
%     \begin{tabular}{c|cccc}
%         \toprule\hline
%          & HR@50 & NDCG@50 & R@50 & P@50 \\ \midrule
%          [1,1,1] & 39.12 & 10.45 & 1.68 & 1.07 \\
%          [1,2,1] & 40.71 & 10.73 & \textbf{1.73} & \textbf{1.10} \\
%          [2,1,1] & 39.67 & 10.71 & 1.68 & 1.07 \\
%          [1,1,2] & \textbf{40.75} & \textbf{11.07} & 1.72 & 1.10 \\ 
%          [1,1,3] & 39.24 & 10.73 & 1.63 & 1.05 \\
%          [1,2,3] & 39.71 & 11.00 & 1.69 & 1.09 \\
%          \bottomrule
%     \end{tabular}
%     \label{tab:online_perf}
% \end{table}

%% file: conclusion.tex
\section{Conclusions and Future Work}
\label{sec:conclusion}

We believe our work opens a new direction of research in multi-stage IR systems, which needs to take both the selection bias in multiple stages of the system and users' underlying interest into consideration. We proved the effectiveness of this solution framework in both offline experiments and online A/B test. 
There are several important future directions which worth indepth exploration. First, we have taken initial efforts aim to decipher the behavior of multi-stage systems, and a more comprehensive understanding will help us design more efficient and effective ranking algorithms. Second, our focus has been primarily on the alignment between our learning objective and GPRP, thus calling for the need to examine the generalization performance throughout the entire learning processes. Third, it is worthwhile to further customize the general framework for specific stages of the system for better performance. Lastly, exploring optimal solutions to handle multiple or all system stages, rather than just one, remains imperative.

%% file: appendix.tex
\section{Appendix}
\subsection{Proof of Proposition \ref{prop: max min representation}}
\label{appendix: proof min max}
\begin{proof}
    Suppose $\mathcal{I}^{i-1} = \{v_1, v_2, \dots, v_{c_{i-1}}\}$ and $r(u, v_1) \geq r(u, v_2) \geq \dots \geq r(u, v_{c_{i-1}})$ without loss of generality. 
    According to equation (\ref{eq: multi-stage prp}), it is easy to see $\mathcal{I}^{i, PRP} = \textbf{Topk}(\mathcal{I}^{i-1}, r(u, v), c_i) = \{v_i | i \in [c_i]\}$.

    Denote a general $\mathcal{I}^i$ as $\{v_{j_1}, v_{j_2}, \dots, v_{j_{c_i}}\}$, where $j_1 < j_2 < \cdots < j_{c_i}$ and $\{j_k | k \in [c_i]\} \subset [c_{i-1}]$. Then for equation (\ref{eq: max min prp}), it is easy to see:
    \begin{align}
        & \argmax_{\mathcal{I}^{i} \subset \mathcal{I}^{i-1}} \min_{\{v_k \in \mathcal{I}^{i} |k \in [c_4]\}} \sum_{k=1}^{c_4} r(u, v_k) \\
        = & \argmax_{\mathcal{I}^{i} \subset \mathcal{I}^{i-1}} \sum_{k=1}^{c_4} r(u, v_{j_{c_i+1-k}}) \\
        = & \{v_i | i \in [c_i]\} \\
        = & \mathcal{I}^{i, PRP}
    \end{align}
\end{proof}

\subsection{Proof of Proposition \ref{prop: worst case result}}
\label{appendix: proof worst case}
\begin{proof}
    We construct a concrete example to prove this proposition. Without loss of generality, suppose we are at stage $3$, and the final stage is $4$. Let $c_2=4, c_3=3, c_4=1$. Suppose candidate items are $(v_1, p^4_1=0.1, r_1 = 1.0), (v_2, p^4_2=0.2, r_2 = 1.0), (v_3, p^4_3=0.5, r_3 = \epsilon), (v_4, p^4_4=0.0, r_4 = 0.0)$, where $\epsilon$ is a very small constant. Suppose the strategy at stage $i$ is to output the item with highest $p^4$. Then according to PRP, we should choose $\{v_1, v_2, v_3\}$ as the output of stage 3 and it is easy to see corresponding utility is $\epsilon$. While according to GPRP, we should choose $\{v_1, v_2, v_4\}$ as the output with utility 1. Thus the gap between them is $1-\epsilon$. When $\epsilon$ approaches 0, the gap approaches $c_4 = 1$.
\end{proof}

\subsection{Proof of Theorem \ref{thr: main}}
\label{appendix: proof theorem main}
\begin{proof}
    Without loss of generality, suppose $L(u, v_1) \leq L(u, v_2)$, now we only need to prove the inequality $w^{i+1, 4}(u, v_1) \leq w^{i+1,4}(u, v_2)$ holds. Since $L(u, v_1) < L(u, v_2)$, there are two possibilities:
    \begin{itemize}
        \item[1]  In the case $S_1 = S_2$ and $Y_1  \leq Y_2$, since these two items are in the same stage and have been exposed, we know $r(u, v_1) < r(u, v_2)$ and $p^{i+1, 4}(u, v_1) = p^{i+1,4}(u, v_2)$, which are obtained after taking expectation over equations $Y_1 < Y_2$ and $O^{i+1, 4}_1 = O^{i+1, 4}_2$. Therefore, we have $w^{i+1, 4}(u, v_1) < w^{i+1,4}(u, v_2)$.
        \item[2]  In the case $S_1 < S_2$. According to Assumption \ref{assum: two}, we have $\frac{p^{i+1,4}(u, v_2)}{p^{i+1,4}(u, v_1)} \geq a \geq \max \{\frac{r(u, v_1)}{r(u, v_2)}, \frac{r(u, v_2)}{r(u, v_1)}\} \geq \frac{r(u, v_1)}{r(u, v_2)} \geq \frac{1}{a}$. Therefore, $p^{i+1, 4}(u, v_1)r(u, v_1) \leq p^{i+1, 4}(u, v_2)r(u, v_2)$, which is exactly $w^{i+1, 4}(u, v_1) \leq w^{i+1,4}(u, v_2)$.
    \end{itemize}
    % Thus we prove the first part of this theorem.
    
    % When $z_0 = z_1 = \dots = z_4 = 0$ and $z_5 = 1$, this is the algorithm we mentioned at the beginning of Section \ref{sec:method}, and the conclusion holds because $\mathbf{E}[L(u, v)] = \mathbf{E}[O^{i+1, 4}(u, v)Y(u, v)] = w^{i+1, 4}(u, v)$.

    Now we finish the whole proof.
\end{proof}

\subsection{Implementation Details of Prerank / Rank Model in Simulated Multi-Stage Pipeline} 
\label{appendix: pipeline}
The settings of prerank / rank model in simulated multi-stage pipeline is listed in Table \ref{tab:simulated_hyperparam}. The label in the multi-stage pipeline equals 1 when: play\_duration >= video\_duration if video\_duration <= 7000, or play\_duration > 7000 if video\_duration > 7000.
\begin{table}[h]
    \centering
    \caption{Settings of prerank / rank in simulated pipeline.}
    \label{tab:simulated_hyperparam}
    \begin{tabular}{c|c|c}
    \toprule \hline
         & prerank & rank \\
    \midrule
       model  & User DNN \& Item DNN & DNN \\ \hline
       hidden layers & [256, 128] & [1024, 512, 256] \\ \hline
       optimizer & \multicolumn{2}{c}{Adam} \\ \hline
       learning rate & \multicolumn{2}{c}{0.001} \\ \hline
       batch size & \multicolumn{2}{c}{8192} \\ \hline
    \bottomrule
    \end{tabular}
\end{table}

In a static request, the following steps are executed in sequence:
\begin{itemize}
    \item[\textbf{Step 1}] For each user, 1,000 items are randomly selected from the prerank candidate pools as the \textbf{simulated prerank candidates}.
    \item[\textbf{Step 2}] For each user, the top 200 items with highest prerank scores in simulated prerank candidates are selected as the \textbf{simulated rank candidates}, while the bottom 200 items with lowest scores in simulated prerank candidates are regarded as the \textbf{simulated prerank negative samples}.
    \item[\textbf{Step 3}] For each user,  the top 50 items with highest rank scores in simulated rank candidates are selected as the \textbf{simulated exposed candidates}, while the bottom 50 items with lowest scores in simulated rank candidates are regarded as the \textbf{simulated rank negative samples}.
    \item[\textbf{Step 4}] For each user, items with positive labels are regarded as the \textbf{simulated exposed positive samples}, while items with negative labels are regarded as the \textbf{simulated exposed negative samples}.
\end{itemize}

\subsection{Data Preparation for Training and Evaluation}
\label{appendix: data preparation} 
Most of the recommendation models are training on the exposed samples, e.g., the retrieval models are trained on the exposed positive samples with randomly sampled negative samples, the rank models are trained on the exposed samples. In our simulated multi-stage recommendation pipeline, models should be trained on the simulated exposed samples. From this perspective, the prerank / rank model in the simulated multi-stage pipeline is not aligned with the real settings. However, it is acceptable since we do not mind the simulated rank model is more powerful in the simulated multi-stage pipeline. The simulated exposed samples are partially randomly generated from the simulated multi-stage pipeline, which may cause difficulty in splitting the dataset for training, validation, and testing. In order to avoid label leakage, we follow the following steps to generate simulated training data:
\begin{itemize}
    \item[\textbf{Step 1}] Generate a static request based on the simulated prerank / rank model and the original prerank candidate pool. The simulated exposed positive samples are reserved as positive samples in the test set, and removed from the original prerank candidate pool.
    \item[\textbf{Step 2}] Generate a static request based on the simulated prerank / rank model and the prerank candidate pool without the positive samples in the test set. The newly simulated exposed positive samples are reserved as positive samples in the validation set, and removed from the prerank candidate pool.
    \item[\textbf{Step 3}] Generate multiple static requests based on the simulated prerank / rank model and the prerank candidate pool without the positive samples in the validation and test set. Samples of each request can be utilized as training samples.
\end{itemize}

Complete separation of the train, validation, and test set is accomplished by the steps above. In our request-based training setting, the training request is set to a random request. Exposed samples and multi-stage samples are randomly sampled from the randomly selected request to build multiple pair-wise or single list-wise optimization objectives.

%% file: reference.bib
@String{Computing = "Computing" }

@String{Computer = "{IEEE} Computer" }

@String{Springer = "Springer-Verlag" }

@ArtifactSoftware{R,
    title = {R: A Language and Environment for Statistical Computing},
    author = {{R Core Team}},
    organization = {R Foundation for Statistical Computing},
    address = {Vienna, Austria},
    year = {2019},
    url = {https://www.R-project.org/},
}

@article{robertson1977probability,
  title={The probability ranking principle in IR},
  author={Robertson, Stephen E},
  journal={Journal of documentation},
  volume={33},
  number={4},
  pages={294--304},
  year={1977},
  publisher={MCB UP Ltd}
}

@article{wechsler2000probability,
  title={The probability ranking principle revisited},
  author={Wechsler, Martin and Sch{\"a}uble, Peter},
  journal={Information Retrieval},
  volume={3},
  pages={217--227},
  year={2000},
  publisher={Springer}
}

@inproceedings{youtubeDNN,
title	= {Deep Neural Networks for YouTube Recommendations},
author	= {Paul Covington and Jay Adams and Emre Sargin},
year	= {2016},
booktitle	= {Proceedings of the 10th ACM Conference on Recommender Systems},
address	= {New York, NY, USA}
}

@inproceedings{MIND_model,
author = {Li, Chao and Liu, Zhiyuan and Wu, Mengmeng and Xu, Yuchi and Zhao, Huan and Huang, Pipei and Kang, Guoliang and Chen, Qiwei and Li, Wei and Lee, Dik Lun},
title = {Multi-Interest Network with Dynamic Routing for Recommendation at Tmall},
year = {2019},
isbn = {9781450369763},
publisher = {Association for Computing Machinery},
address = {New York, NY, USA},
url = {https://doi.org/10.1145/3357384.3357814},
doi = {10.1145/3357384.3357814},
booktitle = {Proceedings of the 28th ACM International Conference on Information and Knowledge Management},
pages = {2615–2623},
numpages = {9},
location = {Beijing, China},
series = {CIKM '19}
}

@inproceedings{SASREC,
  title={Self-attentive sequential recommendation},
  author={Kang, Wang-Cheng and McAuley, Julian},
  booktitle={2018 IEEE International Conference on Data Mining (ICDM)},
  pages={197--206},
  year={2018},
  organization={IEEE}
}

@inproceedings{comirec,
  title = {Controllable Multi-Interest Framework for Recommendation},
  author = {Cen, Yukuo and Zhang, Jianwei and Zou, Xu and Zhou, Chang and Yang, Hongxia and Tang, Jie},
  booktitle = {Proceedings of the 26th ACM SIGKDD International Conference on Knowledge Discovery and Data Mining},
  year = {2020},
  pages = {2942–2951},
  publisher = {ACM},
}

@inproceedings{TDM,
author = {Zhu, Han and Li, Xiang and Zhang, Pengye and Li, Guozheng and He, Jie and Li, Han and Gai, Kun},
title = {Learning Tree-Based Deep Model for Recommender Systems},
year = {2018},
isbn = {9781450355520},
publisher = {Association for Computing Machinery},
address = {New York, NY, USA},
url = {https://doi.org/10.1145/3219819.3219826},
doi = {10.1145/3219819.3219826},
booktitle = {Proceedings of the 24th ACM SIGKDD International Conference on Knowledge Discovery \& Data Mining},
pages = {1079–1088},
numpages = {10},
location = {London, United Kingdom},
series = {KDD '18}
}

@inproceedings{chen2019top,
  title={Top-k off-policy correction for a REINFORCE recommender system},
  author={Chen, Minmin and Beutel, Alex and Covington, Paul and Jain, Sagar and Belletti, Francois and Chi, Ed H},
  booktitle={Proceedings of the Twelfth ACM International Conference on Web Search and Data Mining},
  pages={456--464},
  year={2019}
}

@inproceedings{zhou2021contrastive,
  title={Contrastive learning for debiased candidate generation in large-scale recommender systems},
  author={Zhou, Chang and Ma, Jianxin and Zhang, Jianwei and Zhou, Jingren and Yang, Hongxia},
  booktitle={Proceedings of the 27th ACM SIGKDD Conference on Knowledge Discovery \& Data Mining},
  pages={3985--3995},
  year={2021}
}

@inproceedings{joachims2017unbiased,
  title={Unbiased learning-to-rank with biased feedback},
  author={Joachims, Thorsten and Swaminathan, Adith and Schnabel, Tobias},
  booktitle={Proceedings of the tenth ACM international conference on web search and data mining},
  pages={781--789},
  year={2017}
}

@inproceedings{hu2019unbiased,
  title={Unbiased lambdamart: an unbiased pairwise learning-to-rank algorithm},
  author={Hu, Ziniu and Wang, Yang and Peng, Qu and Li, Hang},
  booktitle={The World Wide Web Conference},
  pages={2830--2836},
  year={2019}
}

@inproceedings{saito2020unbiased,
  title={Unbiased pairwise learning from biased implicit feedback},
  author={Saito, Yuta},
  booktitle={Proceedings of the 2020 ACM SIGIR on International Conference on Theory of Information Retrieval},
  pages={5--12},
  year={2020}
}

@article{liu2009learning,
  title={Learning to rank for information retrieval},
  author={Liu, Tie-Yan and others},
  journal={Foundations and Trends{\textregistered} in Information Retrieval},
  volume={3},
  number={3},
  pages={225--331},
  year={2009},
  publisher={Now Publishers, Inc.}
}

@inproceedings{gao2022kuairec,
  author = {Gao, Chongming and Li, Shijun and Lei, Wenqiang and Chen, Jiawei and Li, Biao and Jiang, Peng and He, Xiangnan and Mao, Jiaxin and Chua, Tat-Seng},
  title = {KuaiRec: A Fully-Observed Dataset and Insights for Evaluating Recommender Systems},
  booktitle = {Proceedings of the 31st ACM International Conference on Information \& Knowledge Management},
  series = {CIKM '22},
  location = {Atlanta, GA, USA},
  url = {https://doi.org/10.1145/3511808.3557220},
  doi = {10.1145/3511808.3557220},
  numpages = {11},
  year = {2022},
  pages = {540–550}
}

@article{burges2006learning,
  title={Learning to rank with nonsmooth cost functions},
  author={Burges, Christopher and Ragno, Robert and Le, Quoc},
  journal={Advances in neural information processing systems},
  volume={19},
  year={2006}
}

@article{burges2010ranknet,
  title={From ranknet to lambdarank to lambdamart: An overview},
  author={Burges, Christopher JC},
  journal={Learning},
  volume={11},
  number={23-581},
  pages={81},
  year={2010}
}

@article{koren2009matrix,
  title={Matrix factorization techniques for recommender systems},
  author={Koren, Yehuda and Bell, Robert and Volinsky, Chris},
  journal={Computer},
  volume={42},
  number={8},
  pages={30--37},
  year={2009},
  publisher={IEEE}
}

@article{rendle2012bpr,
  title={BPR: Bayesian personalized ranking from implicit feedback},
  author={Rendle, Steffen and Freudenthaler, Christoph and Gantner, Zeno and Schmidt-Thieme, Lars},
  journal={arXiv preprint arXiv:1205.2618},
  year={2012}
}

@article{jean2014using,
  title={On using very large target vocabulary for neural machine translation},
  author={Jean, S{\'e}bastien and Cho, Kyunghyun and Memisevic, Roland and Bengio, Yoshua},
  journal={arXiv preprint arXiv:1412.2007},
  year={2014}
}

@inproceedings{covington2016deep,
  title={Deep neural networks for youtube recommendations},
  author={Covington, Paul and Adams, Jay and Sargin, Emre},
  booktitle={Proceedings of the 10th ACM conference on recommender systems},
  pages={191--198},
  year={2016}
}

@inproceedings{ie2019slateq,
  title={SLATEQ: a tractable decomposition for reinforcement learning with recommendation sets},
  author={Ie, Eugene and Jain, Vihan and Wang, Jing and Narvekar, Sanmit and Agarwal, Ritesh and Wu, Rui and Cheng, Heng-Tze and Chandra, Tushar and Boutilier, Craig},
  booktitle={Proceedings of the 28th International Joint Conference on Artificial Intelligence},
  pages={2592--2599},
  year={2019}
}

@article{wang2020cold,
  title={Cold: Towards the next generation of pre-ranking system},
  author={Wang, Zhe and Zhao, Liqin and Jiang, Biye and Zhou, Guorui and Zhu, Xiaoqiang and Gai, Kun},
  journal={arXiv preprint arXiv:2007.16122},
  year={2020}
}

@inproceedings{zhou2018deep,
  title={Deep interest network for click-through rate prediction},
  author={Zhou, Guorui and Zhu, Xiaoqiang and Song, Chenru and Fan, Ying and Zhu, Han and Ma, Xiao and Yan, Yanghui and Jin, Junqi and Li, Han and Gai, Kun},
  booktitle={Proceedings of the 24th ACM SIGKDD international conference on knowledge discovery \& data mining},
  pages={1059--1068},
  year={2018}
}

@inproceedings{zhou2019deep,
  title={Deep interest evolution network for click-through rate prediction},
  author={Zhou, Guorui and Mou, Na and Fan, Ying and Pi, Qi and Bian, Weijie and Zhou, Chang and Zhu, Xiaoqiang and Gai, Kun},
  booktitle={Proceedings of the AAAI conference on artificial intelligence},
  volume={33},
  number={01},
  pages={5941--5948},
  year={2019}
}

@inproceedings{chang2023pepnet,
  title={Pepnet: Parameter and embedding personalized network for infusing with personalized prior information},
  author={Chang, Jianxin and Zhang, Chenbin and Hui, Yiqun and Leng, Dewei and Niu, Yanan and Song, Yang and Gai, Kun},
  booktitle={Proceedings of the 29th ACM SIGKDD Conference on Knowledge Discovery and Data Mining},
  pages={3795--3804},
  year={2023}
}

@article{chang2023twin,
  title={TWIN: TWo-stage Interest Network for Lifelong User Behavior Modeling in CTR Prediction at Kuaishou},
  author={Chang, Jianxin and Zhang, Chenbin and Fu, Zhiyi and Zang, Xiaoxue and Guan, Lin and Lu, Jing and Hui, Yiqun and Leng, Dewei and Niu, Yanan and Song, Yang and others},
  journal={arXiv preprint arXiv:2302.02352},
  year={2023}
}

@inproceedings{liu2017cascade,
  title={Cascade ranking for operational e-commerce search},
  author={Liu, Shichen and Xiao, Fei and Ou, Wenwu and Si, Luo},
  booktitle={Proceedings of the 23rd ACM SIGKDD International Conference on Knowledge Discovery and Data Mining},
  pages={1557--1565},
  year={2017}
}

@inproceedings{pi2020search,
  title={Search-based user interest modeling with lifelong sequential behavior data for click-through rate prediction},
  author={Pi, Qi and Zhou, Guorui and Zhang, Yujing and Wang, Zhe and Ren, Lejian and Fan, Ying and Zhu, Xiaoqiang and Gai, Kun},
  booktitle={Proceedings of the 29th ACM International Conference on Information \& Knowledge Management},
  pages={2685--2692},
  year={2020}
}

@article{feng2021grn,
  title={GRN: Generative Rerank Network for Context-wise Recommendation},
  author={Feng, Yufei and Hu, Binbin and Gong, Yu and Sun, Fei and Liu, Qingwen and Ou, Wenwu},
  journal={arXiv preprint arXiv:2104.00860},
  year={2021}
}

@inproceedings{pei2019personalized,
  title={Personalized re-ranking for recommendation},
  author={Pei, Changhua and Zhang, Yi and Zhang, Yongfeng and Sun, Fei and Lin, Xiao and Sun, Hanxiao and Wu, Jian and Jiang, Peng and Ge, Junfeng and Ou, Wenwu and others},
  booktitle={Proceedings of the 13th ACM conference on recommender systems},
  pages={3--11},
  year={2019}
}

@inproceedings{jia2021pairrank,
  title={Pairrank: Online pairwise learning to rank by divide-and-conquer},
  author={Jia, Yiling and Wang, Huazheng and Guo, Stephen and Wang, Hongning},
  booktitle={Proceedings of the Web Conference 2021},
  pages={146--157},
  year={2021}
}

@article{rubin1974estimating,
  title={Estimating causal effects of treatments in randomized and nonrandomized studies.},
  author={Rubin, Donald B},
  journal={Journal of educational Psychology},
  volume={66},
  number={5},
  pages={688},
  year={1974},
  publisher={American Psychological Association}
}

@article{robertson2009probabilistic,
  title={The probabilistic relevance framework: BM25 and beyond},
  author={Robertson, Stephen and Zaragoza, Hugo and others},
  journal={Foundations and Trends{\textregistered} in Information Retrieval},
  volume={3},
  number={4},
  pages={333--389},
  year={2009},
  publisher={Now Publishers, Inc.}
}

@inproceedings{gey1994inferring,
  title={Inferring probability of relevance using the method of logistic regression},
  author={Gey, Fredric C},
  booktitle={SIGIR’94: Proceedings of the Seventeenth Annual International ACM-SIGIR Conference on Research and Development in Information Retrieval, organised by Dublin City University},
  pages={222--231},
  year={1994},
  organization={Springer}
}

@article{singhal2001modern,
  title={Modern information retrieval: A brief overview},
  author={Singhal, Amit and others},
  journal={IEEE Data Eng. Bull.},
  volume={24},
  number={4},
  pages={35--43},
  year={2001}
}

@inproceedings{yin2016ranking,
  title={Ranking relevance in yahoo search},
  author={Yin, Dawei and Hu, Yuening and Tang, Jiliang and Daly, Tim and Zhou, Mianwei and Ouyang, Hua and Chen, Jianhui and Kang, Changsung and Deng, Hongbo and Nobata, Chikashi and others},
  booktitle={Proceedings of the 22nd ACM SIGKDD International Conference on Knowledge Discovery and Data Mining},
  pages={323--332},
  year={2016}
}

@inproceedings{craswell2008experimental,
  title={An experimental comparison of click position-bias models},
  author={Craswell, Nick and Zoeter, Onno and Taylor, Michael and Ramsey, Bill},
  booktitle={Proceedings of the 2008 international conference on web search and data mining},
  pages={87--94},
  year={2008}
}

@inproceedings{joachims2017accurately,
  title={Accurately interpreting clickthrough data as implicit feedback},
  author={Joachims, Thorsten and Granka, Laura and Pan, Bing and Hembrooke, Helene and Gay, Geri},
  booktitle={Acm Sigir Forum},
  volume={51},
  number={1},
  pages={4--11},
  year={2017},
  organization={Acm New York, NY, USA}
}

@article{zhang2023divide,
  title={Divide and Conquer: Towards Better Embedding-based Retrieval for Recommender Systems From a Multi-task Perspective},
  author={Zhang, Yuan and Dong, Xue and Ding, Weijie and Li, Biao and Jiang, Peng and Gai, Kun},
  journal={arXiv preprint arXiv:2302.02657},
  year={2023}
}

@inproceedings{qin2022rankflow,
  title={RankFlow: Joint Optimization of Multi-Stage Cascade Ranking Systems as Flows},
  author={Qin, Jiarui and Zhu, Jiachen and Chen, Bo and Liu, Zhirong and Liu, Weiwen and Tang, Ruiming and Zhang, Rui and Yu, Yong and Zhang, Weinan},
  booktitle={Proceedings of the 45th International ACM SIGIR Conference on Research and Development in Information Retrieval},
  pages={814--824},
  year={2022}
}

@inproceedings{gallagher2019joint,
  title={Joint optimization of cascade ranking models},
  author={Gallagher, Luke and Chen, Ruey-Cheng and Blanco, Roi and Culpepper, J Shane},
  booktitle={Proceedings of the twelfth ACM international conference on web search and data mining},
  pages={15--23},
  year={2019}
}

@inproceedings{oosterhuis2020policy,
  title={Policy-aware unbiased learning to rank for top-k rankings},
  author={Oosterhuis, Harrie and de Rijke, Maarten},
  booktitle={Proceedings of the 43rd International ACM SIGIR Conference on Research and Development in Information Retrieval},
  pages={489--498},
  year={2020}
}

@inproceedings{ovaisi2020correcting,
  title={Correcting for selection bias in learning-to-rank systems},
  author={Ovaisi, Zohreh and Ahsan, Ragib and Zhang, Yifan and Vasilaky, Kathryn and Zheleva, Elena},
  booktitle={Proceedings of The Web Conference 2020},
  pages={1863--1873},
  year={2020}
}

@inproceedings{ovaisi2021propensity,
  title={Propensity-independent bias recovery in offline learning-to-rank systems},
  author={Ovaisi, Zohreh and Vasilaky, Kathryn and Zheleva, Elena},
  booktitle={Proceedings of the 44th International ACM SIGIR Conference on Research and Development in Information Retrieval},
  pages={1763--1767},
  year={2021}
}
